\documentclass[a4paper,preprint]{emulateapj}
\usepackage{amsmath}
\usepackage{amssymb}
\usepackage{mathrsfs}

\begin{document}

\title{The Case for a Misaligned Relativistic Jet from SN 2001\lowercase{em}}

\author{Jonathan Granot and Enrico Ramirez-Ruiz\footnote{Chandra
Fellow}} \affil{Institute for Advanced Study, Einstein Drive,
Princeton, NJ 08540}


\begin{abstract}

SN 2001em, identified as a Type Ic supernova, has recently been
detected in the radio and X-rays, $\gtrsim 2\;$yr after the
explosion. The high luminosities at such late times might arise
from a relativistic jet viewed substantially off-axis that becomes
visible only when it turns mildly relativistic and its emission is
no longer strongly beamed away from us. Alternatively, the
emission might originate from the interaction of the SN shell with
the circumstellar medium. We find that the latter scenario is hard
to reconcile with the observed rapid rise in the radio flux and
optically thin spectrum, $F_\nu\propto\nu^{-0.36\pm 0.16}t^{1.9\pm
0.4}$, while these features arise naturally from a misaligned
relativistic jet. The high X-ray luminosity provides an
independent and more robust constraint -- it requires $\sim
10^{51}\;$erg in mildly relativistic ejecta. The source should
therefore currently have a large angular size ($\sim 2\;$mas)
which could be resolved in the radio with VLBA. It is also
expected to be bipolar and is thus likely to exhibit a large
degree of linear polarization ($\sim 10\%-20\%$). The presence of
a relativistic outflow in SN 2001em would have interesting
implications. It would suggest that several percent of SNe Ib/c
produce mildly relativistic jets, with an initial Lorentz factor
$\Gamma_0\gtrsim 2$, while the fraction that produce GRB jets
(with $\Gamma_0\gtrsim 100$) is $\sim 100$ times smaller. This
could considerably increase the expected number of transients
similar to orphan GRB afterglows in the radio, and to a lesser
extent in the optical and X-rays, if there is a continuous
distribution in $\Gamma_0$. Furthermore, this may give further
credence to the idea that core collapse SNe, and in particular SNe
Type Ib/c, are triggered by bipolar jets.

\end{abstract}


\section{Introduction}

Supernova (SN) 2001em was discovered on September 15, 2001 in the
nearby galaxy UGC 11794 \citep{P01}, at a redshift of
$z=0.019493$. This corresponds to a distance of $D\approx 80\;$Mpc
(for $\Omega_\Lambda=0.7$, $\Omega_M=0.3$ and $h=0.71$). It was
classified as a Type Ib/c SN \citep[most likely Ic,][]{FC01}. SNe
Type Ib/c -- some of which are
thought to arise from the core collapse of a
Wolf-Rayet (WR) star -- have drawn more attention in recent years
due to their association with gamma-ray bursts (GRBs). The best
and most secure association so far is between GRB 030329 and SN
2003dh \citep{S03,H03}. A compelling case also exists for SN
1998bw (at $z=0.0085$) and GRB 980425 \citep{Ga98}.

This raised interest in the search for signatures of GRB jets in
nearby Type Ib/c SNe \citep[e.g.,][]{Pa01}. Typically, the narrow GRB
jets point away from us and will not be detectable in $\gamma$-rays,
but the SN might still be observed. As the off-axis GRB jets become
mildly relativistic, months to years after the explosion, their
radiation is no longer strongly beamed away from us, and they could
become detectable in the radio.

Thus motivated, \citet{S04} observed a large sample of SNe Ib/c at
late times, and detected SN 2001em on October 17.18, 2003 at
$8.4\;$GHz as a $1.151 \pm 0.051\;$mJy radio source. In addition
to its high radio luminosity, $L_R \sim 10^{28}\;{\rm erg\;
s^{-1}\;Hz^{-1}}$ \citep[second only to SN 1998bw;][]{K98}, SN
2001em was also unusual in its subsequent evolution. Its
$8.4\;$GHz flux rapidly increased to $1.480 \pm 0.052\;$mJy on
January 30.90, 2004. This corresponds to a temporal index of
$\alpha=1.9\pm 0.4$, where $F_\nu\propto\nu^\beta t^\alpha$.
Interestingly, the source appeared nonthermal, exhibiting a
spectral slope of $\beta=-0.36\pm 0.16$ between 4.9 and
$14.9\;$GHz, at the second epoch. More recently, on April 4.81
2004, Chandra detected SN 2001em in the X-ray ($0.5-8\;$keV) with
a luminosity of $L_X\sim 10^{41}\;{\rm
  erg\;s^{-1}}$ and $\beta\approx -0.1\pm0.35$
\citep{Pooley04}.

In this Letter we investigate different explanations for the
unusual emission from SN 2001em. The two most natural mechanisms
are (i) the interaction between the SN shell and the circumstellar
medium (CSM), and (ii) off-axis relativistic jets. We examine
these two possibilities in detail in \S \ref{SN} and \S \ref{GRB},
respectively. Our conclusions are discussed in \S \ref{diss}.

\section{Interaction between the Supernova Shell and the Circumstellar Medium}
\label{SN}

The characteristic SN radio lightcurves are thought to arise from
the competing effects of a slowly declining nonthermal radio
emission and a more rapidly declining absorption. Under the
assumption that the fractions of internal energy in magnetic
fields ($\epsilon_B$) and in relativistic electrons ($\epsilon_e$)
remain constant with time, the observed radio flux can, to first
approximation, be written as $F_\nu\propto\nu^\beta t^\alpha
e^{-\tau}$, where $\beta=\frac{1-p}{2}$ and $p$ is the power-law
index of the electron energy spectrum \citep{C94}. The early
optically thick phase, $\tau \gtrsim 1$, can be dominated by
either free-free absorption or synchrotron self-absorption. The
high expansion velocities and low CSM densities found in type Ib/c
SNe suggest that synchrotron self-absorption is the dominant
mechanism in these objects \citep{C98}. Synchrotron
self-absorption leads to a power law in both time and frequency,
$F_\nu \propto \nu^{5/2}$, instead of an exponential form for
free-free absorption.

SN 2001em showed both a fast rise in its radio flux, and an
optically thin spectral slope, $F_\nu \propto \nu^{-0.36\pm
0.16}t^{1.9 \pm 0.4}$. While the former may be similar to that
expected from synchrotron self-absorption, the latter is clearly
not. The fact that the rapid rise occurs with little absorption,
implies that it is not because of a reduction in optical depth.
The usual models described above therefore fail to reproduce the
observed increase in flux. This behavior has not been observed
previously in radio SNe, although SN 1987A (Type II) has shown a
strong rise in its radio flux \citep{Ball95} together with an
optically thin spectral slope, $\beta\approx-0.95$ \citep{M02},
that has been attributed to interaction with the dense wind from a
previous evolutionary phase \citep{C92}.

In order to address the question of whether or not the radio
emission seen in SN 2001em is consistent with synchrotron
radiation from the interaction of the SN shell with the CSM, we
generalize the analysis of \citet{Waxman04a}, which applies to
expansion in a $\rho_{\rm ext} \propto r^{-2}$ medium, to
$\rho_{\rm ext}=A r^{-k}$. Let us consider a sub-relativistic
shell ejected by the SN explosion, with mass $M$, total energy
$E$, and initial velocity $v_0$. Denoting, $t_{\rm dec}$, the time
at which the SN shell decelerates significantly, we have $t_{\rm
dec}=[2(3-k)E/4\pi Av_0^{5-k}]^{1/(3-k)}$ and $v\approx
v_0\times\min[1,(t/t_{\rm dec})^{(k-3)/(5-k)}]$.
The sharp rise, $\alpha=1.9\pm 0.4$, and the spectral slope,
$\beta=-0.36\pm 16$, that were observed in SN 2001em, cannot be
achieved after $t_{\rm dec}$ \citep{FWK00}. On the other hand, at
$t \ll t_{\rm dec}$, the observed spectral slope suggests that we
are in the power law segment of the spectrum where
$\beta=\frac{1-p}{2}$, which implies $\alpha=3-\frac{k(5+p)}{4}$.
In order to obtain $\alpha\approx 1.9$ one needs $k\lesssim
0.55-0.63$ for $2<p<3$. Such a smooth power law density profile is
unlikely in the immediate surroundings
of a massive star \citep{GS96}.

Explaining the X-ray luminosity, $L_X\sim 10^{41}\;{\rm
erg\;s^{-1}}$ at $t\approx 950\;$days, is not trivial. We have
$L_X\sim f_X\epsilon_{\rm rad}\epsilon_e(E/t)\min[1,(t/t_{\rm
dec})^{3-k}]$ where $f_X$ is the fraction of the radiated energy
in the $0.5-8\;$keV Chandra range, $\epsilon_{\rm
rad}\approx\min[1,(\gamma_m/\gamma_c)^{p-2}]$ is the fraction of
the energy in electrons that is radiated away, and
$\min[1,(t/t_{\rm dec})^{3-k}]$ is the fraction of the total
energy $E$ that is in the shocked CSM.
This implies $(3f_X)(10\epsilon_{\rm
rad})(3\epsilon_e)E_{51}\min[1,(t_{\rm dec}/10^3\;{\rm
d})^{k-3}]\sim 1$, where $E_{51}=E/(10^{51}\;{\rm erg})$, which
suggests that\footnote{The bare minimum for the energy content is
$E\sim 10^{49}\;$erg for $f_X\epsilon_{\rm rad}\epsilon_e=1$. Such
an extreme efficiency is, however, highly unlikely. For more
reasonable values of $f_X,\epsilon_e\sim 1/3$ and $\epsilon_{\rm
rad}\sim 0.1$, we need $E\sim 10^{51}\;$erg.} $E_{51}\gtrsim 1$
and $t_{\rm dec}\lesssim 10^3\;$days. The latter condition implies
$v_0/c\gtrsim 0.5(E_{51}/A_*)^{1/3}$ for $k=2$, where
$A_*=A/(5\times 10^{11}\;{\rm gr\;cm^{-1}})$. As a consequence,
the velocity of the ejecta must be at least mildly relativistic
with $E_{51}\sim 1$.\footnote{In this case, only a small part of
the mass in SN shell, $M\sim E/c^2\sim 5\times
10^{-4}E_{51}\;M_\odot$, would have an initial velocity $v_0\sim
c$.}

The extrapolated radio flux in $8.4\;$GHz at the time of the X-ray
observation is $\sim 1.7\;$mJy, which corresponds to a radio
luminosity of $L_R\sim 10^{38}\;{\rm erg\;s^{-1}}$. This would
lead to $\beta\approx -0.6$ for a single power law in that energy
range, which is consistent with $p\approx 2.25$, as long as
$\nu_c\gtrsim 10^{18}\;$Hz. 
The ratio $L_R/L_X$ requires $p\lesssim 2.25$, where $p<2.25$
gives $\nu_c<10^{18}\;$Hz and $\nu_c (p=2) \sim 10^{16}\;$Hz. Such
high values of $\nu_c$ favor a low CSM density, $A_*\lesssim
0.03(3\epsilon_B)^{-1}(1+Y)^{-4/3}$, where $Y$
is the Compton y-parameter which satisfies
$Y(1+Y)\sim(v/c)\epsilon_{\rm rad}\epsilon_e/\epsilon_B$.
Interestingly, a similarly low value of $A_*$ is required in order
to explain the lack of detection of an off-axis GRB jet in SN
1998bw
\citep{Waxman04a,Waxman04b,SFW04}.\footnote{\citet{Waxman04b} also
derived $v_0\sim 0.8c$ for SN 1998bw, although with a relatively
low energy of $E\sim 10^{49}\;$erg.}

\section{Emission from an Off-Axis Relativistic Jet}
\label{GRB}

We first consider the off-axis emission from a uniform double
sided jet with an initial half-opening angle $\theta_0$ and sharp
edges \citep[e.g.,][]{G02}. Later, we briefly address `structured'
jets, where the energy per solid angle, $\epsilon$, smoothly
decreases with the angle $\theta$ from the jet symmetry axis,
$\epsilon\propto\theta^{-2}$ \citep{R02,ZM02}.

Following \cite{GL03} and generalizing their results to a stellar wind
external density profile, $\rho_{\rm ext}=Ar^{-2}$, we obtain
expressions for the radius $R_j$ where the Lorentz factor $\gamma$ of
the jet drops to $\theta_0^{-1}$, and the radius $R_{\rm NR}$ where
the jet becomes sub-relativistic,
\begin{equation}\label{R_j}
R_j \equiv R_{\rm NR}/f = E/2\pi Ac^2=3.5\times
10^{17}E_{51}A_*^{-1}\;{\rm cm}\ ,
\end{equation}
where  $f\approx 1-\ln\theta_0$, and $E=10^{51}E_{51}\;$erg is the
energy of the jets. The typical angular size of the jet at the
non-relativistic transition time \citep[see][]{GL03}, $t_{\rm
NR}\sim R_{\rm NR}/c$,  is
\begin{equation}\label{theta_NR}
\theta_{\rm NR}=\frac{R_{\rm
NR}}{D_A}=0.71\left(\frac{f}{3}\right)
\frac{E_{51}}{A_*}\left(\frac{D_A}{100\;{\rm
Mpc}}\right)^{-1}\;{\rm mas}\ ,
\end{equation}
where $D_A$ is the angular distance to the source. At the distance
of SN 2001em, $\theta_{\rm NR}=0.88(f/3)E_{51}A_*^{-1}\;$ mas.

The temporal index $\alpha \sim 2$ is consistent with the rising
part of the light curve for a GRB jet viewed off-axis from an
angle of $\theta_{\rm obs}\gtrsim\;{\rm a\ few}\;\theta_0$ w.r.t.
the jet axis \citep[e.g., Fig. 2 of][]{G02}. Therefore, we expect
the peak flux to occur at $t_{\rm peak}=3C\;$yr, where $C\gtrsim
1$. The peak flux at $\nu=8.4\;$GHz should be about $F_{\nu,{\rm
peak}}\approx 2C^2\;$mJy. Given the late peak time, it is likely
that $\theta_{\rm obs}\gtrsim 1$ and therefore $t_{\rm peak}\sim
t_{\rm NR} \sim 3\;$yr, for which the source angular size is
$\sim\theta_{\rm NR}$. According to Eqs. \ref{R_j} \&
\ref{theta_NR}, $\theta_{\rm NR}\sim 2-3\;$mas. One can also
estimate $\theta_{\rm NR}$ by requiring an apparent velocity of
$c$, $\theta_{\rm NR}\sim ct_{\rm NR}/D_A\sim 2.4(t_{\rm
NR}/3\;{\rm yr})\;$mas. Such an angular size could be resolved by
VLBA.

In order to explain the spectral slope of $\beta \sim -0.4$, we
require that $\nu_m<\nu<\nu_c$, for which 
$\beta=\frac{1-p}{2}$. The measured value of $\beta$ can be
somewhat larger than this asymptotic value if $\nu_m\sim 1\;$GHz.
Following \citet{NPG02}
and 
\citet{GS02}, we find
\begin{eqnarray}
F_{\nu,{\rm peak}} &=& 285\frac{g(p)}{g(2.2)}a^{-p}
\epsilon_{e,-1}^{p-1}\epsilon_{B,-2}^{(p+1)/4}
A_*^{3(p+1)/4} \nonumber
\\ \label{F_peak2}
& & \times E_{51}^{(1-p)/2} \nu_{10}^{(1-p)/2}\theta_{\rm
obs}^{-2p}\;{\rm mJy}\ ,
\\ \label{t_peak}
t_{\rm peak} &=& a\left(\frac{\theta_{\rm
obs}}{\theta_0}\right)^2t_j = 34(1+z)a\frac{E_{51}}{A_*}\theta_{\rm
obs}^2\;{\rm days}\ ,
\end{eqnarray}
(for $\theta_{\rm obs}\gtrsim 2\theta_0$ and at the redshift of SN
2001em) where
$g(p)=(p-0.18)e^{-1.66p}\left(\frac{p-2}{p-1}\right)^{p-1}$, and
$a$ relates $t_j$ to $t_{\rm peak}$. For $\theta_{\rm obs}\sim 1$
we expect $a\sim 4$, while for $\theta_{\rm obs}\ll 1$ we expect
$a\sim 1$.
For SN 2001em, $F_{\nu,{\rm peak}}(10\;{\rm
GHz})\approx 2C^2\;$mJy, which equals the flux in Eq.
\ref{F_peak2} for $a\approx 4$ and $\theta_{\rm obs}\approx
C^{-1/p}(\pi/2)$.
This suggests a viewing angle $\theta_{\rm obs}\gtrsim
C^{-1/p}\;$rad. Since for SN 2001em we know that $t_{\rm
peak}=3C\;$yr, Eq. \ref{t_peak} yields $aE_{51}A_*^{-1}\theta_{\rm
obs}^2\sim 30C$, which implies a small CSM density, $A_*\sim 0.1$,
similarly to \S \ref{SN}. This relation can also be used to
simplify Eq. \ref{F_peak2} and eliminate the dependence on
$\theta_{\rm obs}$ and $a$,
\begin{equation}\label{F_peak3}
F_{\nu,{\rm peak}}\sim 0.2C^{-p}
\epsilon_{e,-1}^{p-1}\epsilon_{B,-2}^{(p+1)/4}
 A_*^{(3-p)/4}E_{51}^{(p+1)/2}
\nu_{10}^{(1-p)/2}\;{\rm mJy}\ .
\end{equation}
Thus we obtain that
$\epsilon_{e,-1}^{p-1}\epsilon_{B,-2}^{(p+1)/4}
A_*^{(3-p)/4}E_{51}^{(p+1)/2}\sim 10C^p$.  
Assuming a typical energy in GRB jets of $E_{\rm 51}\sim 1$, and
$A_*\sim 0.1$, this gives 
$\epsilon_{e,-1}^{p-1}\epsilon_{B,-2}^{(p+1)/4}\sim 6C^p$. As
discussed in \S \ref{SN}, the ratio $L_X/L_R$ implies $p\lesssim
2.2$. For $C\approx 1$ the above condition can be readily
satisfied for a wide range of reasonable parameter values (e.g.,
$\epsilon_e\sim 0.3$, $\epsilon_B\sim 0.1$). However, since
$\epsilon_e$, $\epsilon_B\lesssim 1/3-1/2$, we must have
$C\lesssim 2-3$, which implies $t_{\rm peak}\lesssim 5-8\;$yr.

Finally, we briefly address a `sructured' GRB jet viewed from a
large angle $\theta_{\rm obs}$. If the jet has an outer edge at
$\theta_{\rm max}<\theta_{\rm obs}$, then the light curve would
not be very different from that for a uniform jet viewed at
$\theta_{\rm obs}>\theta_0$ \citep[e.g.,][]{WJ03}. In this case,
the above analysis is still approximately valid. If, on the other
hand, $\theta_{\rm max}=\pi/2$ or $\theta_{\rm obs}<\theta_{\rm
max}$, then the early light curve is dominated by emission from
material along the line of sight. In this case, a sharp rise like
the one observed in SN 2001em ($\alpha\approx 2$), together with
the observed spectral slope, $\beta\sim -0.4$, cannot be achieved
after the time $t_{\rm dec}$ when the material along the line of
sight decelerates significantly \citep{GS02,KG02,GK03}. Therefore,
the only way this scenario might still work is if we are before
$t_{\rm dec}$.  In this case $t_{\rm peak}\gtrsim 3\;$yr is given
by $t_{\rm dec}\sim t_{\rm NR}\Gamma_0^{-2(4-k)/(3-k)}$. This
suggests a mildly relativistic initial Lorentz factor along the
line of sight, $\Gamma_0\lesssim{\rm a\ few}$, which might also
explain why no GRB or X-ray flash was observed
\citep[e.g.,][]{RL02}, despite the very low redshift of SN 2001em.
Similarly to the non-relativistic case discussed in \S \ref{SN},
$\alpha\sim 2$ requires $k\lesssim 0.6$, which is unlikely.

\section{Discussion}
\label{diss}

Different possible explanations for the radio emission from SN
2001em $\gtrsim 2\;$yr after the SN have been considered. We find
that the large temporal index, $\alpha=1.9\pm 0.4$, together with
the optically thin spectral slope, $\beta=-0.36\pm 0.16$, cannot
be naturally explained as emission from the interaction between
the SN shell and the CSM. This would require either an almost
uniform external density, or a density bump
\citep[e.g.,][]{RDMT01}. On the other hand, we find that a GRB
jet, or even a jet with a mildly relativistic initial Lorentz
factor, $\Gamma_0\gtrsim 2$, that points away from us can
naturally reproduce the observed temporal and spectral properties.

Since the actual observed rise in the radio luminosity was only
$\sim 30\%$, it might still not be indicative of a long episode of
increasing flux and could be only due to a local density bump.
However, the measured X-ray luminosity provides a stronger and
more robust constraint. It requires $\sim 10^{51}\;$erg in ejecta
with a mildly relativistic expansion velocity. Such a system would
be physically very similar to 
an initially relativistic
jet which became mildly relativistic at $t\sim t_{\rm NR}$, and
began to approach spherical symmetry. It is also reasonable to
expect that the mildly relativistic SN ejecta would be somewhat
elongated along the rotational axis,
similar to a relativistic jet near $t_{\rm NR}$. A high degree of
linear polarization might therefore be expected. The polarization
from a relativistic jet viewed off-axis is expected to reach its
maximum value near the time of the peak in the light curve,
$t_{\rm peak}$. For a relativistic jet the peak polarization can
reach $\sim 30\%-40\%$, while for a mildly relativistic jet it is
probably more modest, $\sim 10\%-20\%$, but still significantly
higher than for a typical SN.

The best way to test our conclusion of a mildly relativistic
expansion velocity
is via the angular size of the image, which should be $\gtrsim
2\;$mas, and could be resolved with VLBA.
For a double sided relativistic jet, we might observe both jets,
if the viewing angle is large enough, $\theta_{\rm obs}\gtrsim 1$,
so that the difference in brightness between the two jets would
not be very large \citep{GL03}. In this case, their brightness
ratio and its temporal evolution can help determine our viewing
angle, $\theta_{\rm obs}$.

If indeed the radio and X-ray emission observed in SN 2001em are
from an off-axis relativistic jet, then this has several
interesting implications. This could provide an estimate for the
fraction $f_{\rm RJ}$ of SNe Ib/c that produce relativistic jets.
In order to account for the observed emission, we only need an
initial Lorentz factor of $\Gamma_0\gtrsim 2$. Such jets would
generally not produce a GRB, which typically requires
$\Gamma_0\gtrsim 100$. In this case, if we use a conservative
estimate, combining the 33 SNe from the sample of \citet{B03}, and
the additional 7 (including 2001em) from the sample of
\citet{SFW04}, then SN 2001em would be one out of $40$ nearby SNe
Ib/c that produced relativistic jets. This implies $f_{\rm
RJ}\gtrsim 2.5\%$. Following \citet{SFW04} and using only nearby
SNe Ib/c for which there are late time ($>100\;$days)
observations, we obtain $f_{\rm RJ}\sim \frac{1}{15}\approx
6.7\%$. Since the observations are sparse (and in most cases
consist of a single upper limit) the actual value of $f_{\rm RJ}$
might even be larger.

It is interesting to compare $f_{\rm RJ}$ to the fraction $f_{\rm
GRB}$ of SNe Ib/c that produce GRBs. There are various estimates
for $f_{\rm GRB}$. Assuming a uniform jet with sharp edges,
\citet{F01} found a beaming correction of $\langle
f_b^{-1}\rangle\sim 500$ between the observed and the true GRB
rates (where $f_b\approx\theta_0^2/2$) that results in $f_{\rm
GRB}\approx 0.4\%$. \citet{PSF03} estimated $f_{\rm GRB}$ for the
universal structured jet (USJ) model and found $f_{\rm GRB}\sim
8\times 10^{-6}$. \citet{GPW03} found that the USJ model is not
consistent with the observed $\log N-\log S$ distribution, and did
a more thorough analysis for the uniform jet model, that resulted
in $\langle f_b^{-1}\rangle\approx 75\pm 25$ and $f_{\rm
GRB}\approx(5.5\pm 1.8)\times 10^{-4}$. Therefore, if indeed
$f_{\rm RJ}\gtrsim\;$a few percent, then $f_{\rm RJ}/f_{\rm
GRB}\sim 10^2$, implying that SNe Ib/c produce $\sim 100$ times
more mildly relativistic jets (with $\Gamma_0\gtrsim 2$) than
highly relativistic ones (with $\Gamma_0\gtrsim 100$), as
suggested by several authors \citep[][]{MWH01,RCR02,GL03}.

If this is the case, one might expect a smooth and continuous
distribution $P(\Gamma_0)$ of initial Lorentz factors for the jets
produced by SN Ib/c, where $\Gamma_0\gtrsim 100$ would produce a
GRB, $\Gamma_0\gtrsim 10-20$ could result in X-ray orphan
afterglows and possibly also X-ray flashes, $\Gamma_0\gtrsim 5-10$
may give rise to optical orphan afterglows, and $\Gamma_0\gtrsim
2$ could be responsible for radio orphan afterglows. If, for
example, we parameterize this distribution as a power law,
$P(\Gamma_0)=K\Gamma_0^{-\eta}$ for $\Gamma_{\rm
min}<\Gamma_0<\Gamma_{\rm max}$, where $\Gamma_{\rm min}\approx
1$, $\Gamma_{\rm max}>100$, and\footnote{Here $\int
P(\Gamma_0)d\Gamma_0=f_{\rm RJ}$ is normalized to the total
fraction of SNe Ib/c that produce relativistic jets.} $K=f_{\rm
RJ}(\eta-1)\Gamma_{\rm min}^{1-\eta}$, then we need $\eta \sim 2$
in order to get $f_{\rm RJ}/f_{\rm GRB}\sim 10^2$. However, this
is still highly speculative at this stage.

We now
compare $f_{\rm RJ}/f_{\rm GRB}\sim 10^2$ to current observational
limits.  \citet{L02} have used the FIRST and NVSS surveys to place
limits on orphan radio GRB afterglows. They estimated the number
of candidates for such events over the whole sky above $6\;$mJy to
be 227, and obtained a lower limit on
$f_b^{-1}\sim E_{\gamma,{\rm iso}}/E$, of $\langle
f_b^{-1}\rangle>13$. However, this was derived assuming a fixed
value for the isotropic equivalent energy output in $\gamma$-rays,
$E_{\gamma,{\rm iso}}$, while allowing the true energy $E$ to
vary. If instead we fix the true energy to be $E\approx
10^{51}\;$erg, as suggested by \citet{F01} and \citet{BFK03}, the
same analysis would result in an upper limit 
of $\langle f_b^{-1}\rangle\lesssim 6300$. Following
\citet{GPW03}, $f_{\rm RJ}/f_{\rm GRB}\sim 10^2$ and $\langle
f_b^{-1}\rangle=75\pm 25$, which is consistent with the revised
limit of $\langle f_b^{-1}\rangle\lesssim 63$ that is obtained by
scaling up the expected number of such transients by a factor of
$f_{\rm RJ}/f_{\rm GRB}$. This gives roughly the right number of
radio transients found by \citet{L02}, if most of them are caused
by $\Gamma_0\gtrsim 2$ jets produced in SNe Ib/c.

\citet{NP03} estimated the ratio of on-axis orphan X-ray
afterglows ($\Gamma_0\gtrsim 10-20$) and GRBs ($\Gamma_0\gtrsim
100$) to be less than $8$, using the ROSAT all sky survey. This is
marginally consistent with $\eta\sim 2$ and suggests $\eta\lesssim
2$. Finally, we note that even if the radio emission from SN
2001em arises from the deceleration of a relativistic jet, then
there is still a large statistical uncertainty on the value of
$f_{\rm RJ}$, since it is estimated on the basis of one event. For
example, $f_{\rm RJ}/f_{\rm GRB}$ might still be $\sim 10$, which
would imply $\eta\sim 1.5$.

A relatively large value of $f_{\rm RJ}$ might support the idea
that at least some core collapse SNe, and in particular SNe Ib/c,
may be triggered by bipolar jets \citep[][]{K99}. Even if only
$\sim 1\%$ of the core energy is channelled into such jets, they
would still have enough kinetic energy to provide most of the
power in the explosion, and substantially alter the structure of
the expanding SN shell. While most rotating magnetized
proto-neutron stars with low power are expected to produce broad
slowly collimating jets, a few high power ones should produce
narrow rapidly collimating jets \citep[][]{U92,T94}. Although
carrying more power, these highly collimated jets will be much
less efficient than the broad jets in imparting energy and
momentum to the outer layers \citep{K99}. They may then act
similarly to the failed SNe \citep{MWH01,I04}, continuing to
accrete much of the surrounding stellar layers and collapse to a
black hole, potentially resulting in even faster and narrower
jets. Observational estimates of the ratio $f_{\rm RJ}/f_{\rm
GRB}$ will be valuable for constraining the different stellar
evolution routes involved in producing relativistic, bipolar jets
in core collapse SNe.

\acknowledgments

We thank B. Paczy\'nski and S.~D. Van Dyk for helpful discussions
that initiated this work.
This work is supported by the W.M. Keck
foundation, NSF grant PHY-0070928 (JG), and NASA through a Chandra
Postdoctoral Fellowship award PF3-40028 (ER-R).

\end{document}